\documentclass[nofootinbib,twocolumn,prc,aps]{revtex4-1}
\usepackage[utf8]{inputenc}
\usepackage{amsmath,bm}

\usepackage{graphicx,srcltx,amsmath,color}

\def\eps{\epsilon}
\def\pa{\partial}

\newcommand{\beq}{\begin{equation}}
\newcommand{\eeq}{\end{equation}}
\newcommand{\beqa}{\begin{eqnarray}}
\newcommand{\eeqa}{\end{eqnarray}}
\newcommand{\apgt} {\ {\raise-.5ex\hbox{$\buildrel>\over\sim$}}\ }

\newcommand{\fract}[2]{{\textstyle\frac{#1}{#2}}}
\newcommand{\fracd}[2]{{\displaystyle\frac{#1}{#2}}}

\newcommand{\vb}[1]{\mbox{\boldmath $#1$}} %

\newcommand{\dP}{{\partial \cal P}}

\newcommand{\cC}{{\cal C}}
\newcommand{\dC}{{\partial \cal C}}

\newcommand{\dxc}{d^3\!x}
\newcommand{\vbx}{\vb{x}}
\newcommand{\dd}[1]{\delta^{#1}}

\graphicspath{{figs/}}

\begin{document}

\title{The role of the electric Bond number in the stability of   pasta phases}

\author{Sebastian Kubis}
\email{skubis@pk.edu.pl}
\author{W\l{}odzimierz~W\'ojcik}
\affiliation{Institute of Physics, Cracow University of Technology,   Podchor\c{a}żych 1, 
30-084 Kraków, Poland}

\begin{abstract}

The stability of pasta phases in cylindrical and spherical Wigner--Seitz (W--S) cells is examined. The electric Bond number is 
introduced as the ratio of electric and surface energies. In the case of a charged rod in vacuum, other kinds of instabilities appear in addition to the well known 
Plateau--Rayleigh mode. For the case of a rod confined in a W--S cell
the variety of unstable modes is reduced. It comes from the virial theorem, which bounds the value of the Bond number from above 
and reduces the role played by electric forces. A similar analysis is done for the spherical W--S cell, where it appears that the  
inclusion of the virial theorem stabilizes all of the modes.

\end{abstract}

\maketitle

\section{Introduction}

In a neutron star, there is a transitional region between the solid crust and the liquid core, where neutron-rich nuclei could be strongly deformed. Ravenhall
{\em et al.} presented the first approach to this kind of structure in \cite{ravenhall83}
based on the liquid drop model of nuclei. It was shown that the competition between the surface and Coulomb energies 
of nuclei immersed in a quasi-free neutron liquid leads to exotic shapes like infinite cylinders or flat layers called
pasta phases.
 In order to describe the two-phase system of a proton cluster immersed in a neutron liquid, Ravenhall \cite{ravenhall83}
used the Wigner--Seitz approximation---an isolated neutral cell with a shape assumed to be a ball, cylinder, or slab.
Comparison of the cell energy of different shapes will show which one of them is preferred. When the volume occupied by the 
proton cluster increases, there is a sequence of phases: balls, cylinders, and slabs, followed by their inversions:
 the cylindrical-hole and the spherical-hole, where the proton phase surrounds the neutron phase. 
The simplicity of this approach encouraged many authors to analyze  pasta phases
with regards to  various  neutron star properties like  their cooling, precession, oscillations  and other transport properties. For a review of this work, see \cite{Schmitt:2017efp}.

{ The stability of pasta phases against shape perturbation is still an open question.
The stability analysis is, to some extent, parallel to the consideration of the elastic properties of the pasta and its 
oscillations. Elastic properties in the liquid drop model framework were first derived by Pethick and Potekhin in 
\cite{Pethick:1998qv} and later, by other authors in \cite{Durel:2018cxs,Pethick:2020aey}. Another approach, i.e., 
molecular dynamics, was presented in \cite{Caplan:2018gkr}. These works have shown the relevance of the interplay between 
the Coulomb and surface energies.
However, when the elastic properties were analyzed, only certain classes of modes, representing the relative displacements of 
slabs or rods in the long-wavelengthlimit, are relevant. A detailed inspection of changes in pasta shape is not 
included. In this work, we focus on pasta shape stability due to the 
competition between the surface and Coulomb energies.}

The presence of surface tension may affect the shape stability of different structures in different ways..
For a neutral system, it is commonly known that a spherical droplet is always stable \cite{barbosa} --- the surface energy goes to the global minimum under the constraint of conserved volume. However, it appears 
that the same surface energy may destabilize the fluid portion in a cylinder, if its 
length is greater than its circumference, $L> 2\pi R$. This is the well-known Plateau--Rayleigh instability \cite{rayleigh}.
Because the pasta supplies the fluid with surface tension and a charge, the question arises of whether the exotic
shapes of nuclear clusters are stable. Some attempts to analyze the stability of the pasta phases were made by 
Pethick \&
Potekhin in \cite{Pethick:1998qv} and Iida {\em et al.} \cite{Iida:2001xy}. In the work by Pethick and Potekhin, the elastic properties 
of pasta phases in the form of periodically placed slabs (lasagna) and rods (spaghetti) 
were considered. It was shown that the elastic constants are positive in both cases, which would guarantee structural 
stability; however, one must be cautious with this conclusion. In the case of lasagna, in order to derive the elastic 
constant,
it was sufficient to consider one specific perturbation mode in the infinite wavelength limit. Complete stability 
analysis should take into account any type of surface perturbation with a finite wavelength. Such analysis was presented 
recently in \cite{Kubis:2017lgv} by inspection of the second-order energy variation for one proton slab in a unit cell 
with periodic boundary conditions. It appeared that the second-order energy variation is positive for all modes and all 
volume 
fractions occupied by the slab in the unit cell. 

Similarly, the same analyses can be carried out for the spaghetti phase. The second-order analysis
of the candidate structure for the minimum of energy must fulfill the necessary condition for the extremum: vanishing of 
the first-order energy variation. Such a condition, the Euler--Lagrange equation for the total energy, is expressed 
by the relation between the mean curvature, $H=\frac{1}{2}(\kappa_1+\kappa_2)$, of the cluster surface (where $\kappa_i$ are 
principal curvatures) and the electric potential~$\Phi$ 
\beq
2\sigma H(\vb{x})= {\cal C} + \Delta\rho\;\Phi(\vb{x})~,
\label{H-eqn}
\eeq 
where $\Delta\rho=\rho_+ \!-\! \rho_-$ is the charge density contrast between phases ($\Delta\rho\! =\! e n_p$, if protons are confined in clusters), $\sigma$ is the surface tension, and $\cal C$ is a constant that depends on the pressure difference between the phases and the potential gauge. For more details, see \cite{Kubis:2016fmw}. In the work of \cite{Pethick:1998qv}, a periodic lattice of rods was considered. In such a system, the potential, $\Phi(x) $, becomes a complicated space-dependent function, so Eq.~(\ref{H-eqn}) indicates that the proton clusters cannot be cylinders with constant curvature $H=\frac{1}{2R}$. The charged cylinder obeys Eq.~(\ref{H-eqn}) only if it is placed in the cylindrical W--S cell or in vacuum. For such a structure, the first-order variation of the energy vanishes, and the examination of the second-order variation can be undertaken. In the work of \cite{Iida:2001xy} the perturbation of a rod in a cylindrical cell
was tested. The authors took only one mode into account, which was the transitionally invariant mode along the axis of the rod, corresponding to the transverse flattening of the rod. In this work, we will complete the stability analysis of a 
single rod in vacuum and in a cylindrical W--S cell by testing any kind of perturbation. Similarly, the stability 
of a charged ball in a spherical W--S cell will also be considered. \\ \\ 
{The structure of the work is as follows: in the section \ref{sec-ggf}, the choice of Green's function is 
discussed, and the generalized Green's function is introduced as the appropriate tool for the derivation of the perturbed potential.}
In the section \ref{sec-rinv}, the stability of a charged rod in vacuum is presented, and the role of the Bond number is emphasized.
In the section \ref{sec-rincell}, the same analysis is done for a rod in a cylindrical W--S cell. The spherical W--S cell is treated in the section \ref{sec-bincell}. 
 
\section{The generalized Green's function}
\label{sec-ggf}
The stability analysis consists of testing the energy change due to the cluster surface deformation 
$\eps(\vb{x})$, where $\vb{x}$describes the position of the unperturbed proton cluster surface. Then, the second order variation of the energy is given by the surface integral over the proton cluster boundary, $\dP$
\begin{equation}
\dd{2}\tilde{E} = \frac{1}{2} \oint_\dP \left(-\sigma (\nabla^2 \eps + B^2 \eps) + 
\Delta\rho \;(\pa_n\Phi\; \eps +\delta\Phi ) \right) \eps \; dS ,
\label{var2nd} 
\end{equation}
where $B^2$ is the sum of the squared principal curvatures \linebreak $B^2=\kappa_1^2+\kappa_2^2$, $\pa_n\Phi$ is the normal derivative of the unperturbed potential, and $\delta\Phi$ is the potential perturbation coming from the change in the charge distribution 
\begin{equation}
 \dd\!\rho(\vb{x}) = \Delta\rho \; \eps \; \delta_\dP(\vb{x}) ,
\label{var1charge}
\end{equation}
where $\delta_\dP(\vb{x})$ is the surface delta function. 
The potential perturbation, $\delta\Phi$, depends linearly on $\eps(\vb{x})$ and is the solution of the Poisson equation
\begin{equation}
  \nabla^2 \delta\Phi(\vbx) = - 4 \pi\; \delta\!\rho(\vbx) ~.
  \label{poisson}
\end{equation}
The potential $\delta\Phi$ depends not only on the charge perturbation $\delta\rho$, but also on the 
boundary conditions assumed for the potential, $\delta\Phi$.
Because we are going to consider an isolated W--S cell,
devoid of periodicity, the boundary condition imposed for Eq.(\ref{poisson}) becomes a matter of discussion.
The W--S cell must be neutral, which means that the electric field flux over the cell boundary $\dC$ is zero
\begin{equation}
   \oint_\dC \vb{E}\cdot\vb{n}\; dS =0 .
   \label{neutrality}
\end{equation}
Therefore, it seems the most natural choice is the Neumann boundary condition 
$\pa_n\delta\Phi|_\dC = 0 $, in order to keep the relation (\ref{neutrality}).
The Dirichlet boundary condition, which specifies the potential value on every point of the cell boundary, is
not appropriate in this case. It is sufficient to assume that the integral of $\pa_n\delta\Phi|_\dC$ over the cell boundary is 
zero. This may be achieved by the introduction of the so-called {\em generalized} Green’s function which shares some of its properties with more common Neumann Green’s function. The Neumann Green’s function fulfils the following equations \cite{jackson}
\beq
\nabla^2 G_N(\vbx,\vbx') = - 4\pi\delta(\vbx-\vbx') ~~{\rm and}~~ \left.\frac{\pa G_N }{\pa n'}\right|_{\vb{x}'\in \pa \cal 
C} = -\frac{4\pi}{S_c} ~,
\eeq 
where $S_c$ is the surface area of the cell boundary, $\dC$.
However, to ensure that the cell is neutral for any charge distribution, we have to use the generalized Green’s function which fulfils another pair of equations \cite{duffy}
 \begin{align}
\nabla^2 G_{\rm gen}(\vbx,\vbx') &=- 4\pi\left(\delta(\vbx-\vbx')-\frac{1}{V_{c}}\right) , \label{Ggen-poisson}\\
 \left. \frac{\pa G_{\rm gen}}{\pa n'}\right|_{\vb{x}'\in \pa \cal C} &= 0 , &  \label{Ggen-bc}
 \end{align}
where $V_c$ is the cell volume.

Now the perturbed potential $\delta\Phi$ is derived by
\begin{equation}
 \delta\Phi(\vbx) = \int_\cC G_{\rm gen}(\vbx,\vbx')\delta\rho(\vbx')\dxc' .
 \label{deltaPhi}
\end{equation}
Green’s function for periodic system fulfills the same 
conditions as Eqs.~(\ref{Ggen-poisson},\ref{Ggen-bc}). For more details, see \cite{marshall}.
Hereafter, the Green's functions used in this work for a W--S cell will be understood to be the generalized
Green's function.

\section{The shape perturbation of a rod in vacuum}
\label{sec-rinv}
Before we proceed with the stability analysis for the \mbox{W--S} cell, we present 
the charged rod in vacuum. Although the case does not correspond exactly to the structure of 
the pasta phases, because of the lack of a neutral cell, it does show an interesting interplay between 
electric and surface forces. Moreover, a comparison of 
the case in vacuum with the case of a W--S cell emphasizes the role played by the boundary conditions.
The formalism presented in the previous section can be applied to this case, with the sole difference of using the vacuum Green’s function which ensures that $\delta\Phi\to 0$ when $r \to\infty$, which means that
\begin{equation}
    G_{\rm vac}(\vbx,\vbx')|_{r'\rightarrow\infty} \rightarrow 0 
\end{equation}
for any points given by $\vbx$. 
The charge contrast is the charge density of the rod $\rho = \Delta\rho$ and the unperturbed potential is 
\begin{eqnarray}
 \Phi(r) =& &   \nonumber \\
-2\pi\rho R^2   \,& &
\begin{cases}
       \fracd{1}{2} \left(\fracd{r^2}{R^2}-1\right)+\ln (R)      & 0<r \le R\\ \\
   \ln r & R<r<\infty ~. \\
 \end{cases}    
\end{eqnarray}
It is sufficient to consider the deformations that are periodic along the $z$-axis, with mode wavelength $L$. Then the Green's function is also periodic in the $z$-coordinate. In the cylindrical coordinates, $r,z,\phi$, the Green's function is represented by the following sum
\begin{align}
 G_{\rm cyl}(\vbx,\vbx') & =  \label{greencyl} \\ \nonumber 
 & \hspace{-3em}  \frac{2}{L}\sum_{n=0}^\infty\sum_{m=0}^\infty \gamma_{nm} g_{nm}(r,r') 
 \cos k_n (z-z')\cos m(\phi-\phi')~,
 \end{align}
where $k_n =2\pi/L$. The coefficient, $\gamma_{nm}$, has the properties $\gamma_{00}=1$ $\gamma_{n0}=\gamma_{0m}=2$   
and $\gamma_{mn}=4$ for $m,n>0$. The key components of this expansion are the $g_{nm}(r,r')$ functions and their form depends on the boundary conditions. In vacuum, they are
\begin{equation}
g_{nm}^{\rm vac}(r,r') = \,
\begin{cases}
 -\ln r_>& n=m=0 \\
 \fracd{1}{2m}\left(\fracd{r_<}{r_{>}}\right)^m & n=0,~ m>0\\
 I_m(k_n r_<) \, K_m(k_n r_>) & n>0,~m>0 ~.
\end{cases}
\end{equation}
The presence of surface tension means that the cylinder of neutral fluid is always unstable. The cylindrical portion of the fluid 
fragmentizes into pieces by because of local narrowing in the shape of an hourglass. The wavelength of the most unstable mode is $L=2\pi R$. Here we show how the situation changes when the cylindrical portion of the fluid in vacuum is charged.
\begin{figure}
   flattening ~~~~~~~~ hourglass   ~~~~~~~~~~~~gyroid\\
   \includegraphics[width=.3\columnwidth]{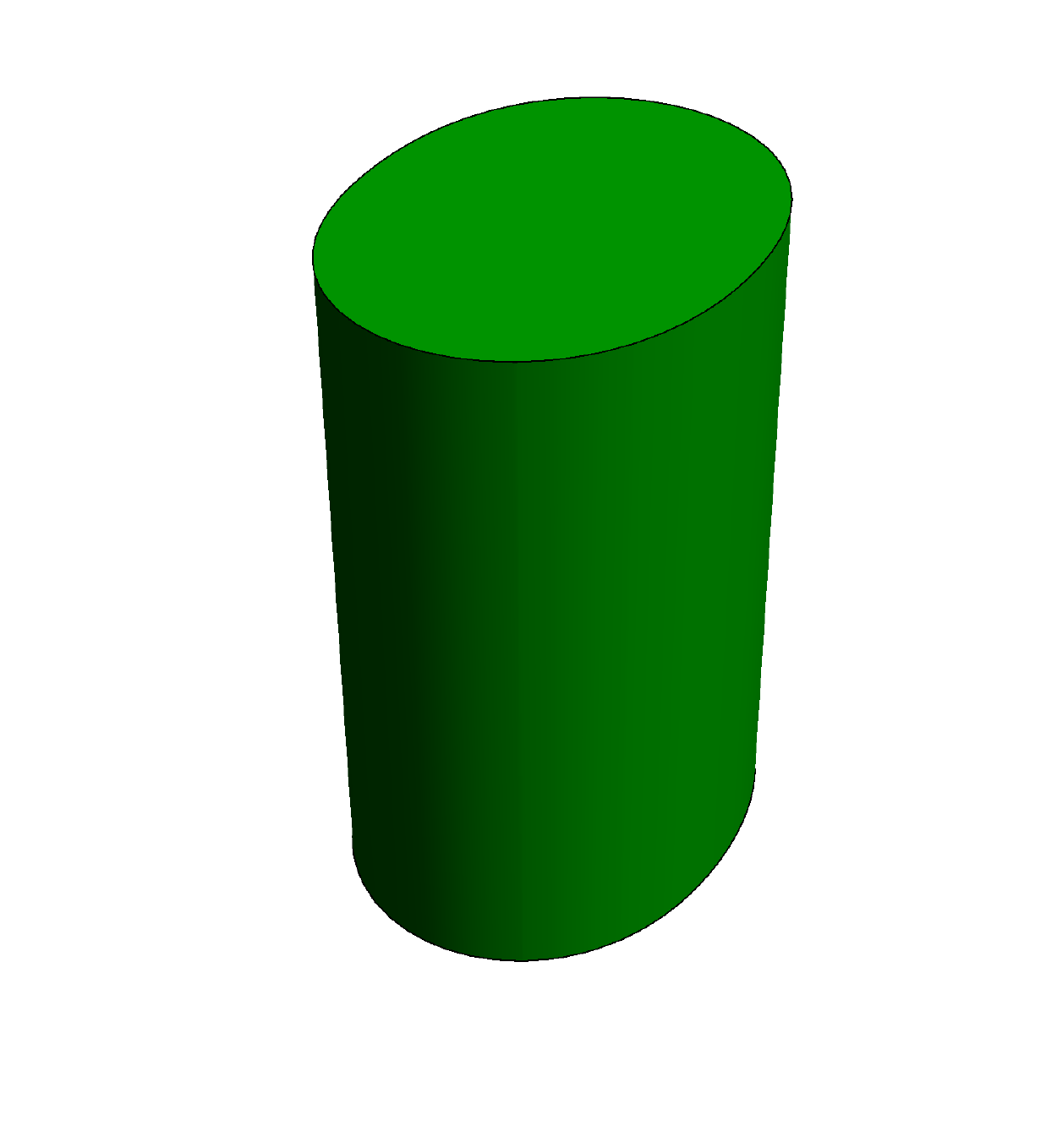}
   \includegraphics[width=.3\columnwidth]{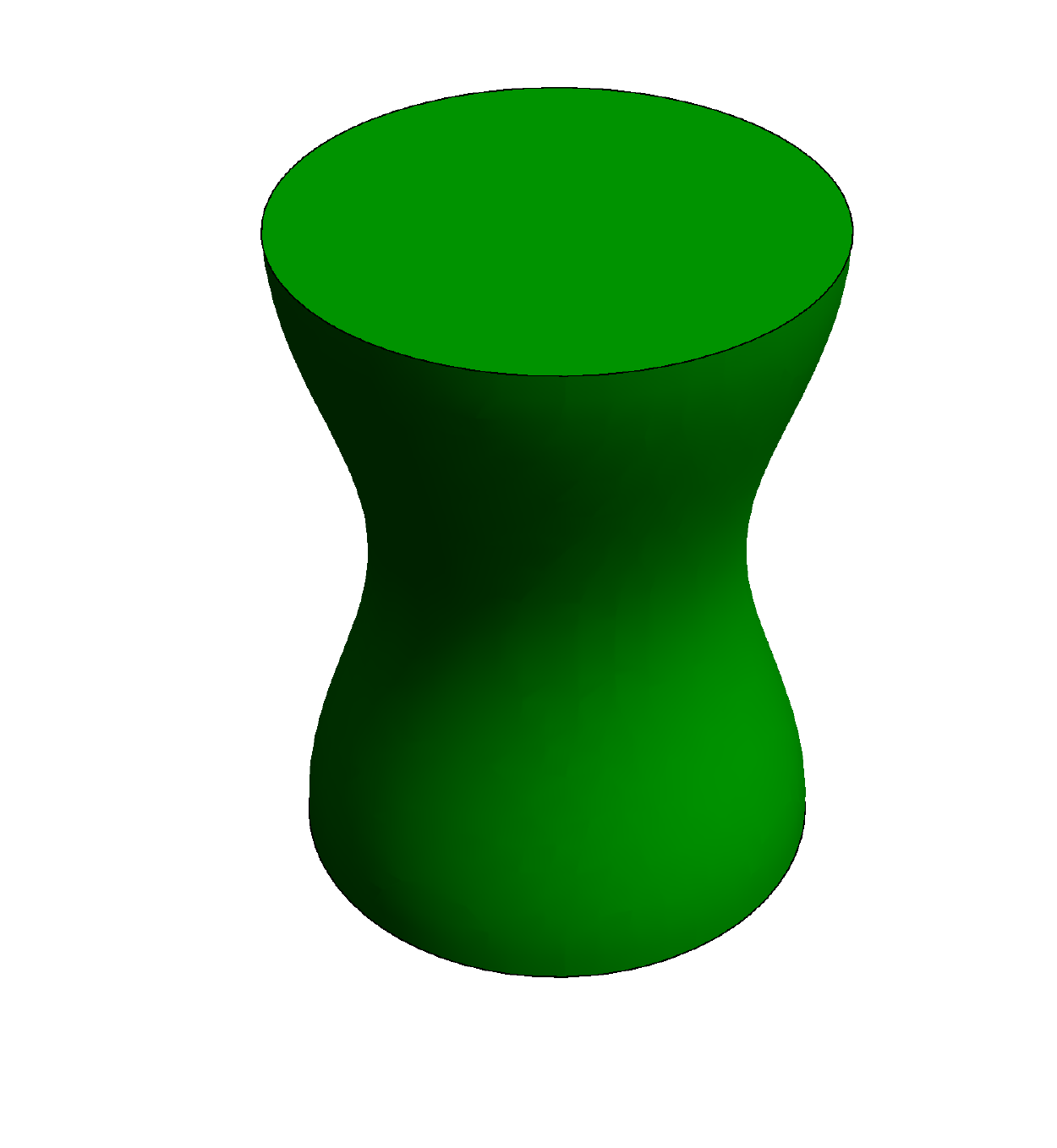}
   \includegraphics[width=.3\columnwidth]{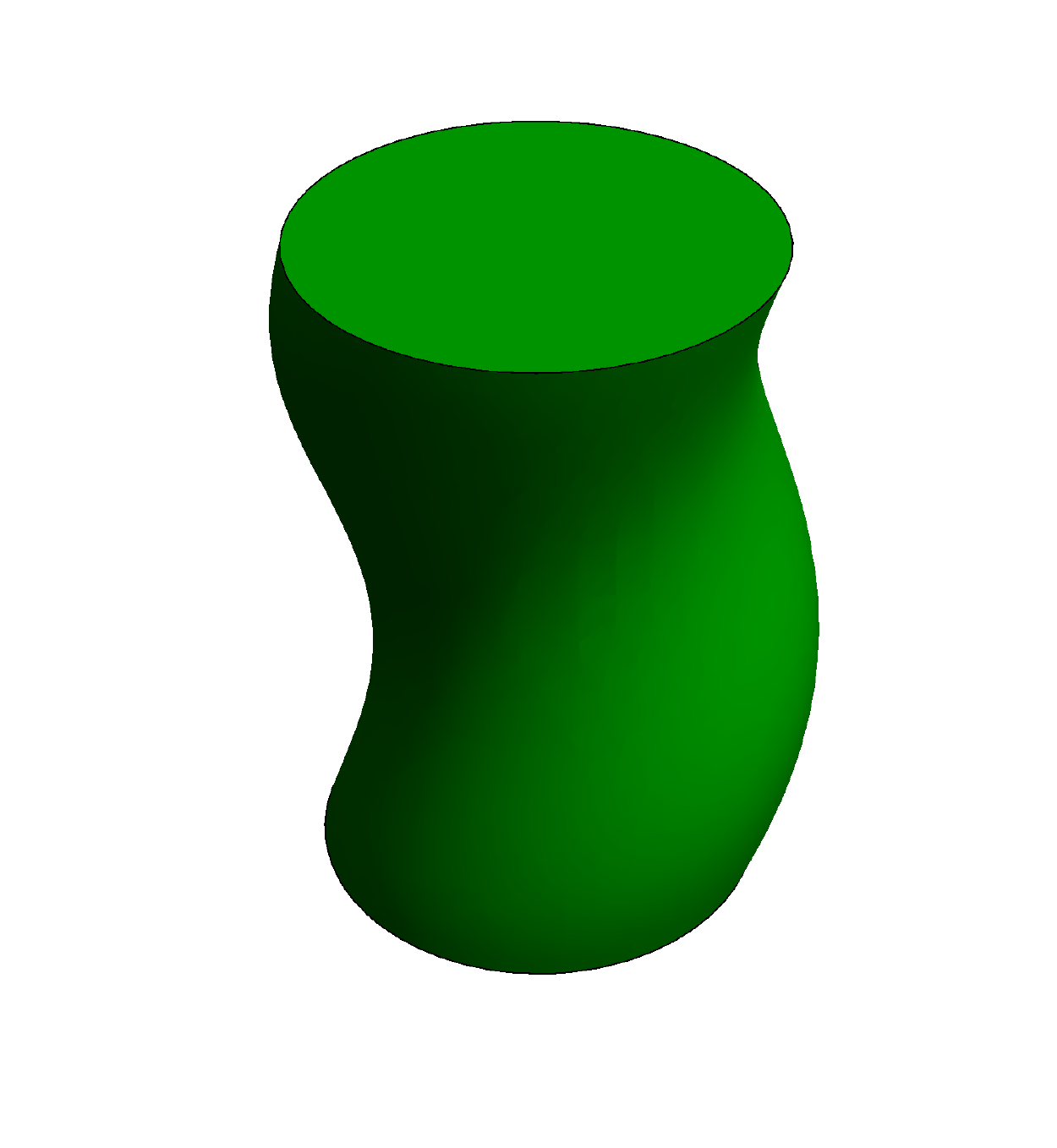}
   \caption{The considered perturbation shapes.}
\label{pert-shapes}
\end{figure}   
It appears that the relevant modes of deformation are:
\begin{equation}
 \eps(z,\phi) = \left\{
\begin{array}{lr}
 \alpha \cos( 2\phi) & \rm flattening \\
 \alpha \cos (2\pi z/L) & \rm hourglass \\
 \alpha \cos(2\pi z/L + \phi)   & \rm gyroid    ~,
 \end{array}
 \right.
\end{equation}
   where $\alpha$ is perturbation amplitude. Their shapes are shown in Fig.~\ref{pert-shapes}. 
 From the derivation of the subsequent terms in Eqs.(\ref{var2nd}), we determine the expression for
 the second variation of the total energy. It is a quadratic function of the deformation amplitude $\alpha$.
 In order to test the sign of the energy variation, it is convenient to express $\delta \tilde{E}$ in terms of a dimensionless function of dimensionless variables:
 \begin{equation}
      f(x,\zeta) = \frac{\delta^2\tilde{E}}{2\pi^3 \alpha^2 R^3 \rho^2}~,
    \label{stabfun} 
 \end{equation}
 where $x=L/2\pi R$ is dimensionless wavelength and
 \begin{equation}
    \zeta = \frac{\pi R^3 \rho^2}{\sigma}
    \end{equation}
is a parameter which measures the ratio of electrostatic forces to surface forces. Such a quantity can be directly related to the well known quantity which appears in the physics of electrofluids, the so-called {\em electric Bond number} \cite{brakke}. It is defined by 
\begin{equation}
   {\rm Bo_e} = \frac{|\vb{E}_0|^2 R}{4 \pi\sigma}~,
   \label{Bo}
\end{equation} 
where $R$ is the characteristic size of a structure, and $\vb{E}_0$ is the typical electric 
field\footnote{Here we use Gaussian units, in SI units
the electric Bond number is expressed as ${\rm Bo_e} =\varepsilon_0|\vec{E}_0|^2 R/\sigma$. } 
If we include the size of the proton cluster and the electric field on its surface
 then the electric Bond number is exactly equal to the $\zeta$ parameter
\begin{equation}
   {\rm Bo_e} = \zeta~. 
\end{equation} 
 The stability function $f$ takes the following form for the three deformations considered:
\begin{multline}
 f^{\rm vac}(x,\zeta) = \\
   \left\{
\begin{array}{ll}
 \fracd{x}{2}\left( \fracd{3}{\zeta }-1 \right) & \rm flattening   \vspace{.5em}   \\
   \fracd{x}{2\zeta}\left(\fracd{1}{x^2} -1 \right) - x+2 x
    I_0(\frac{1}{x}) K_0(\frac{1}{x}) \label{fhour} & \rm hourglass \vspace{.5em} \\
   \fracd{1}{2 x \zeta}-x+2 x I_1(\fract{1}{x}) 
    K_1(\fract{1}{x}) &   \rm gyroid   ~.
\end{array} 
\right.   
\end{multline}
\begin{figure}
   \includegraphics[width=\columnwidth]{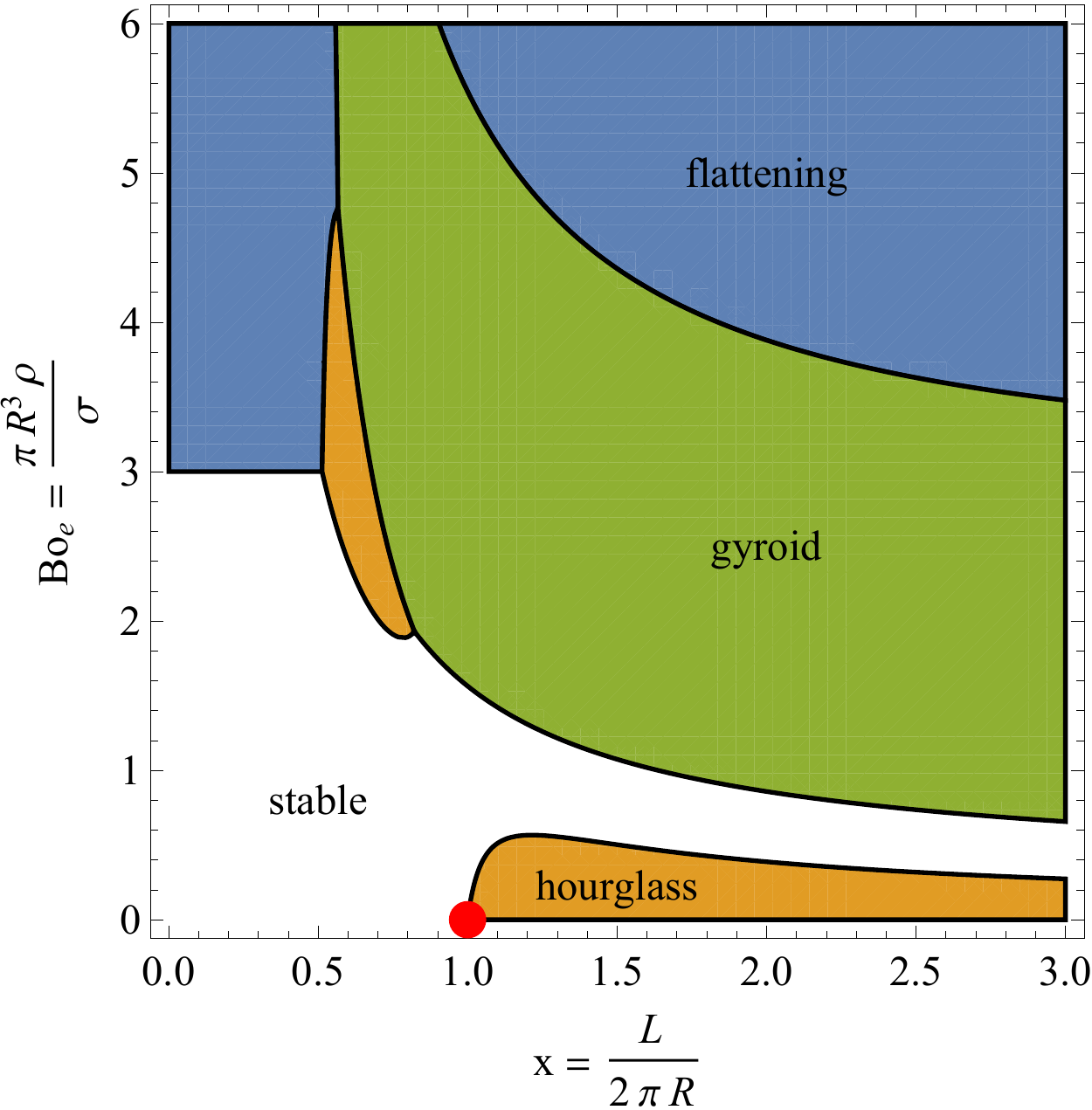}
   \caption{The stability map (color version online). The point on the x-axis represents 
 the Plateau--Rayleigh instability.}
\label{pert-plots}
\end{figure}   
When $f(x,\zeta) <0 $ for a given mode then the mode is unstable and its amplitude may increase.
For all the three modes their stability functions have negative values for some regions in the parameter space
$(x,\zeta)$. For some modes, these regions can overlap; then the most unstable mode is the one with the most negative value for 
$f(x,\zeta)$. The regions for the most unstable modes are depicted indicated in the Fig.~\ref{pert-plots} by their names or colors.
As expected, for large Bond numbers ($\zeta \gg 1$), the most unstable mode is the flattening mode where the charge spreads 
out to diminish the electrostatic energy, which can occur for any mode wavelength.
This kind of instability corresponds to the 
transition from the spaghetti phase to the lasagna phase. For very small Bond numbers, the main instability is represented by the hourglass 
mode, where the surface energy drives the deformation. In the case of an uncharged cluster, i.e., $\zeta \rightarrow 0$, the 
first term in the stability function for the hourglass mode, Eq.~(\ref{fhour}), dominates completely. Then we may recover the 
already mentioned Plateau--Rayleigh 
instability, which is indicated by a point in Fig.~\ref{pert-plots}. However, the most intriguing mode is the gyroidal 
mode, which can be unstable for both the surface energy-dominated region (small $\rm 
Bo_e$) and the charge-dominated region (large $\rm Bo_e$). When the surface energy dominates, the gyroidal mode is unstable 
only if it is sufficiently long. When the electric charge energy dominates, the gyroidal mode is unstable for short wavelength. The 
region of the $(x,\zeta)$-space where the charged cylindrical cluster is stable is represented by the white
color in Fig.{\ref{pert-plot}}. The narrow region for stable modes, located between the hourglass and gyroid instability regions, declines with 
increasing mode wavelength, $x\rightarrow\infty$. In summary, an unexpected and important outcome is that for any value 
of the electric Bond number, there is always at least one type of unstable mode. The charged rod in vacuum is always 
{\em unstable}, but this work shows that the type of unstable mode may be controlled by its charge.

This consideration concerning the single rod may be treated as an approximation to the spaghetti pasta phase when 
the distance between rods is much larger than their perpendicular size. 
The case of a single rod in an isolated Wigner--Seitz cell is presented in the next section.

\section{The shape perturbation of a rod in a W--S cell}
\label{sec-rincell}

Here we consider the cylindrical W--S cell which is electrically neutral as whole and has volume $V_c = \pi^2 R_c L$.
The positively charged rod with $\rho_+ = (1-w)\Delta\rho$ occupies $w$ fraction of the cell volume $V_c$ and is surrounded 
by the negatively charged medium   $\rho_-=- w \Delta\rho$. The volume fraction,   $w$, depends on the cluster radius $w=R^2/R_c^2$.   
For the generalized Green’s function in cylindrical coordinates the expansion given by Eq.~(\ref{greencyl}) is still valid. 
The only change concerns the radial $g$-functions which now have the form:
\begin{figure*}[t!]
   \includegraphics[width=\textwidth]{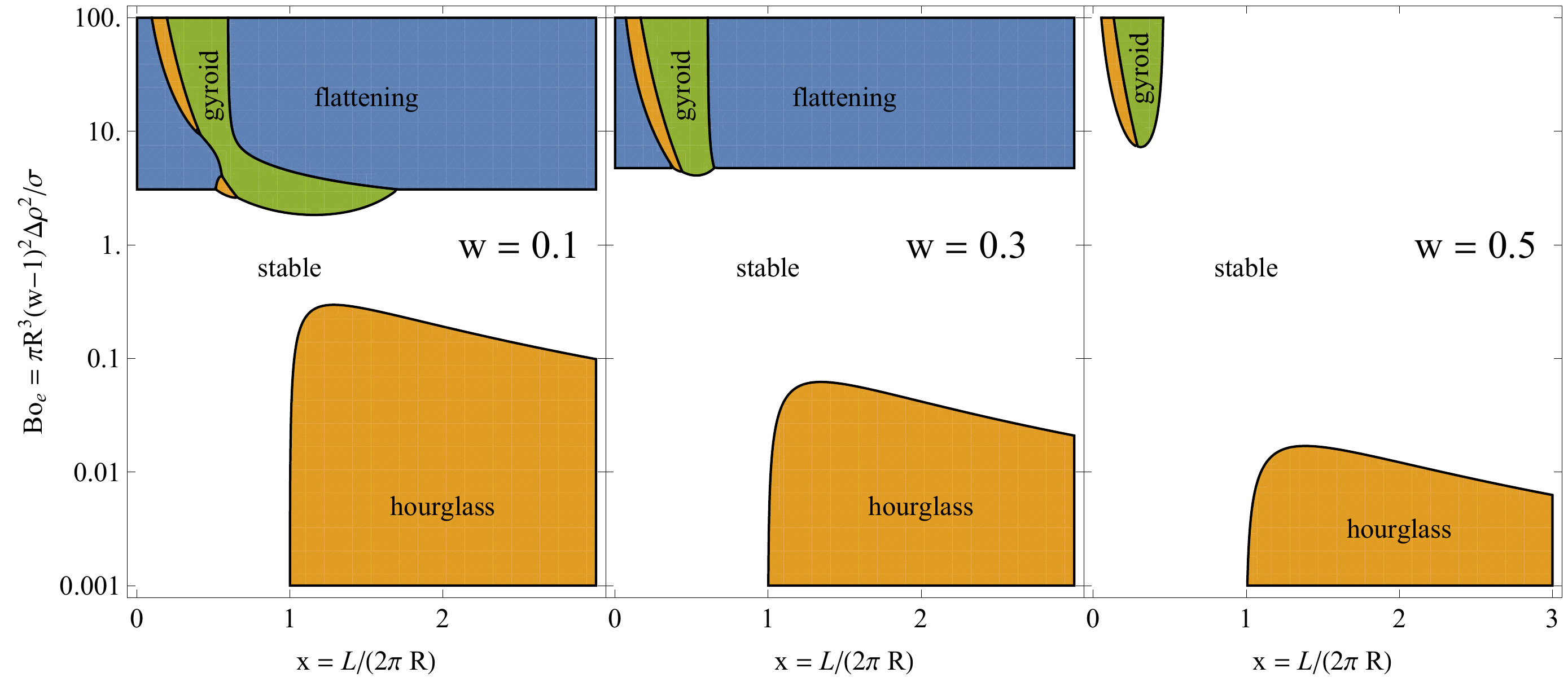}
   \caption{The stability map for a charged rod in a cylindrical W--S cell.
  The colors (online) represent the regions of the ($x,\zeta)$ space where the most
unstable mode occurs. The value of the volume fraction $w$ is shown in the panels.}
   \label{ws-stabmap}
\end{figure*}
\begin{widetext}
\begin{equation}
   g_{nm}^{\rm gen}(r,r') =
   \begin{cases}
\displaystyle
 \frac{r^2+r'^2}{2 R_c^2}-\ln \left(\frac{r_>}{R_c}\right) & ~~ n=m=0 \\   \\
\displaystyle 
   \frac{1}{{2 m}}\left(\frac{r_<}{R_c}\right)^m
    \left(\left(\frac{R_c}{r_>}\right)^m + \left(\frac{r_>}{R_c}\right)^m\right)
    & ~~n=0,~ m>0\\ \\
    \displaystyle
 I_m\left(\fract{2 \pi   n\, r_<}{L}\right) \left(K_m\left(\fract{2 \pi   n\,
    r_>}{L}\right)-\frac{K^\prime_m\left(\frac{2 \pi n R_c}{L}\right)
    I_m\left(\frac{2 \pi n\, r_>}{L}\right)}{I^\prime_m\left(\frac{2 \pi n R_c}
    {L}\right)}\right) &~~ n>0,~m>0 ~~,
\end{cases}
\end{equation}
\end{widetext}
where the W--S cell radius, $R_c=R/\sqrt{w}$ appears explicitly. 

As in the case of the single rod in vacuum, the second variation of the total energy Eq.(\ref{var2nd}) of the cell may be calculated. The stability function $f$, defined by Eq.(\ref{stabfun}), is now a function of three variables: the mode wavelength $x$, the Bond number $\zeta$ and the volume fraction $w$ occupied by the charged rod. The value of the Bond number is still defined by Eq. (\ref{Bo}) but in the W--S cell, the electric field distribution is different than in vacuum, and the Bond number is now
expressed by 
\begin{equation}
 {\rm Bo_e}   = \zeta = \frac{\pi R^3 (w-1)^2 \Delta\rho^2}{\sigma}~.
\end{equation}
For the three types of perturbation mode the   stability functions are given by Eqs.~(\ref{ws-fstab})

\begin{widetext}
\begin{equation}
f^{\rm ws}(x,\zeta,w) = \left\{
\begin{array}{lr}
\fracd{x}{2} (w^2+2 w-1 +\fracd{3}{\zeta}{(w-1)^2} )   & \rm flattenig   \vspace{.5em}   \\
\fracd{2 x I_0(\frac{1}{x})}{ I_1(\frac{1}{x\sqrt{w}})}\left[{I_0(\frac{1}{x})
    K_1(\frac{1}{x\sqrt{w}})} + 
    K_0(\frac{1}{x}) I_1(\frac{1}{x\sqrt{w}}) \right] + x(w - 1)   -
    \fracd{1}{2 \zeta   x} (w-1)^2 (x^2-1)& \rm ~~hourglass \vspace{.5em} \\
2 x I_1(\frac{1}{x})
    \left[\frac{I_1(\frac{1}{x}) (K_0(\frac{1}{x\sqrt{w}})+K_2(\frac{1}{x\sqrt{w}}))}{I_0(\frac{1}{x\sqrt{w}}) + 
    I_2(\frac{1}{x\sqrt{w}})}+K_1(\frac{1}{x})\right]+ x(w-1) +   \fracd{1}{2 \zeta   x} (w-1)^2 &   \rm gyroid ~.
\end{array}
\right.
\label{ws-fstab}
\end{equation}
\end{widetext}
In Fig.~\ref{ws-stabmap}, the regions of instability for the three basic modes are shown in the contour map. For the W--S cell ,the stability functions are functions of the three variables $x,\zeta$, and $w$. In order to compare the 
 results with those in vacuum, we plot the contours of $f(x,\zeta,w)<0$ in $x,\zeta$ space with $w$ fixed.
We have chosen $w=0.1, 0.3$, and $0.5$ and, contrary to the vacuum case, there are Bond numbers where all types of 
modes are stable. The region of stability grows with increasing $w$. We also analyzed larger values of $w$, and when the rod fills the whole cell, $w\to 1$, all unstable regions move towards very large Bond numbers (note the 
logarithmic scale for $\zeta$). The only exception is the hourglass mode, which shrinks to a point $x=1$ in the region of very small~$
\zeta$, corresponding to the Plateau--Rayleigh instability.

\begin{figure}[t!]
\hfill \includegraphics[width=.9\columnwidth]{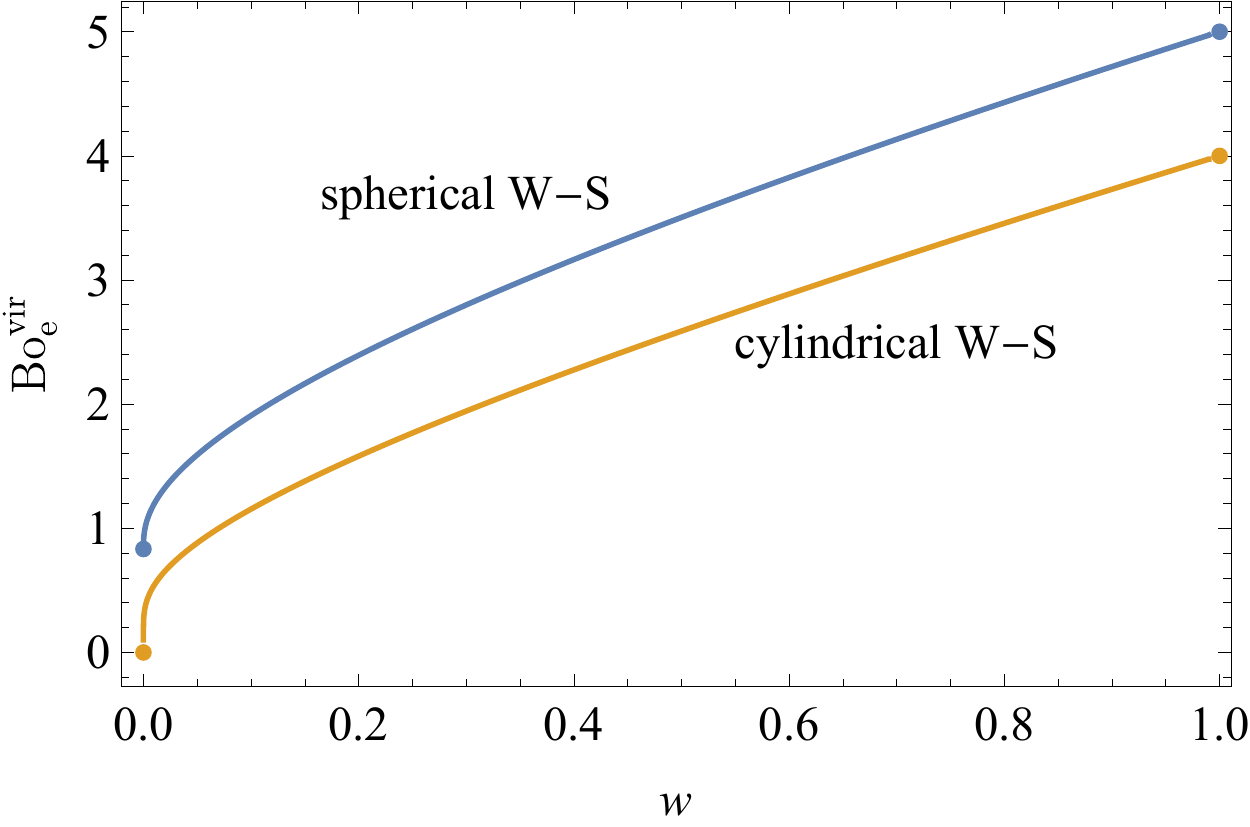}
   \caption{The relation between the Bond number and volume fraction for different W--S cell shapes.}
   \label{Bovir-plot}
\end{figure}

\begin{figure}[h!]
   \includegraphics[width=\columnwidth]{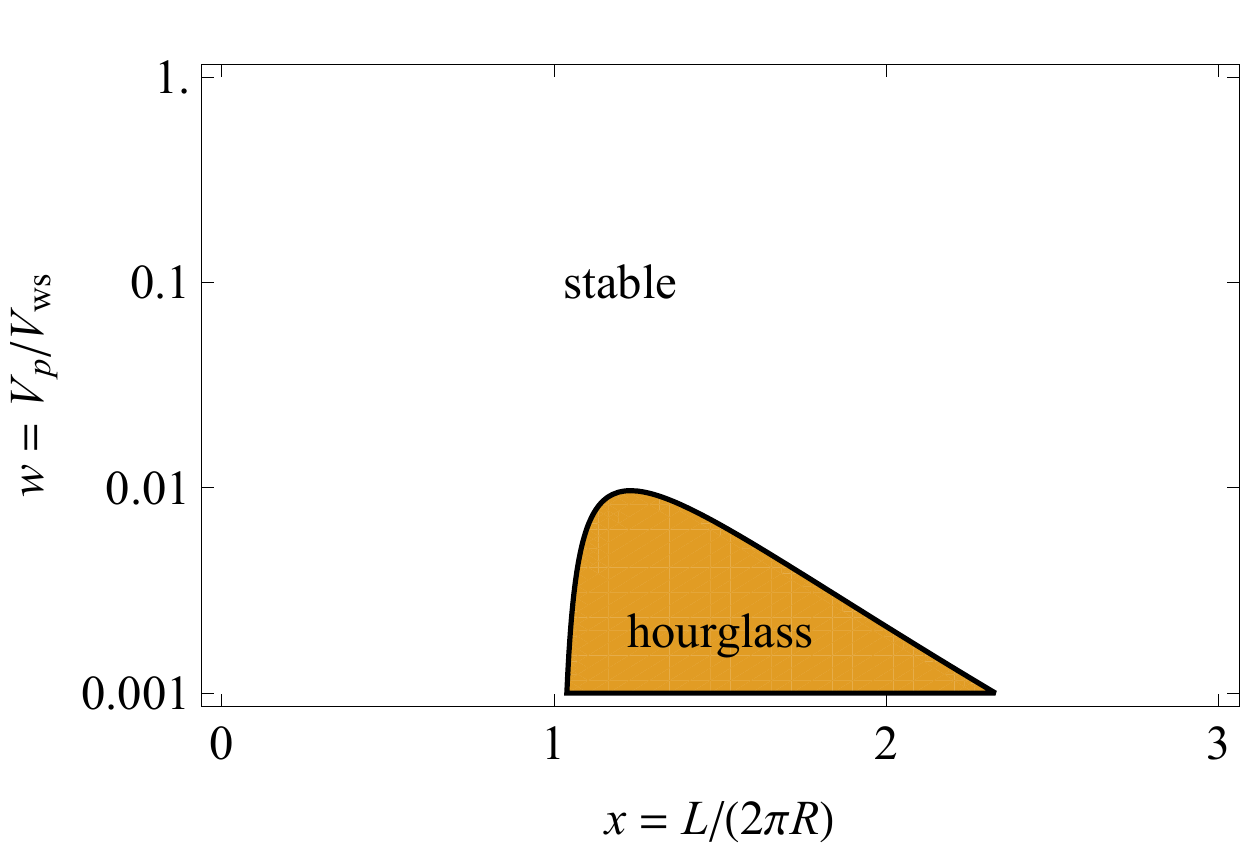}
   \caption{The stability map for the cylindrical W--S cell with the virial theorem included.}
   \label{vir-stabmap}
   \end{figure}

The three variables $x,\zeta$, and $w$ are not independent. The so-called virial theorem says that minimization with respect to the cell size makes the relation between the surface and Coulomb energies $E_s=2E_{Coul}$. This leads to the relation between the charge contrast, $\Delta \rho$, and the surface tension $\sigma$
\begin{equation}
\sigma = \frac{1}{2} \pi   \Delta \rho ^2 R^3 (w-\ln w-1)   ~.   
\end{equation}
Thus, the Bond number in the case of the W--S cell is uniquely determined by the volume fraction $w$
\begin{equation}
{\rm Bo_e^{vir}} = \zeta(w) = \frac{2(w-1)^2}{w - \ln w - 1}   ~.
\label{Bovir}
\end{equation}
The above equation shows that the virial theorem constrains the Bond number. It now ranges between 0 and 4, as it is shown in 
Fig.~\ref{Bovir-plot}. The fact that the Bond number in the W--S cell never goes to a higher value means that the 
electric forces are always smaller or at most comparable to the surface forces. Large values of $\zeta$ are relevant for the instability 
of the flattening and gyroid modes (see Fig.~\ref{pert-plots}). These modes are driven by the spread of charge and may be unstable 
if the electric forces alone are sufficiently large. Thus, when the Bond number is bounded from above, such modes 
become stable inside the W--S cell. The region where $f^{\rm vir}(w,x)<0$ is much smaller (see the logarithmic scale of $w$ in Fig.~\ref{vir-stabmap}) and occurs only for the hourglass mode,
which is shown in Fig.~\ref{vir-stabmap}.

\section{The shape perturbation of a ball in a W--S cell}
\label{sec-bincell}

\begin{figure}[b!]
   \includegraphics[width=\columnwidth]{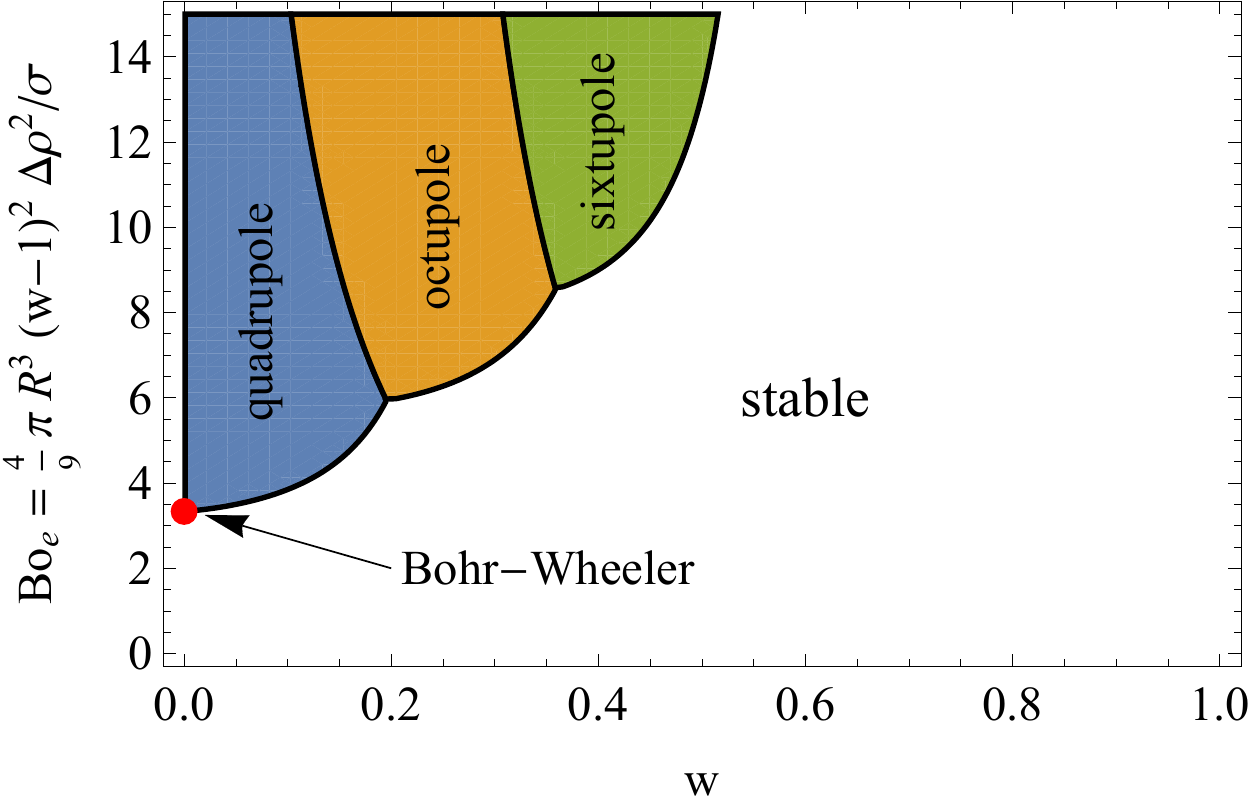}
   \caption{The stability map for a charged ball in a spherical W--S cell.
  The colors (online) represent the regions of the ($w,\zeta)$ space with the most
unstable mode.}
   \label{ws-stabmap-sph}
\end{figure}

The analysis of shape stability for a spherical \mbox{W--S} cell is simpler than for the cylindrical cell. The 
perturbation modes are now expressed by the Legendre polynomials
\begin{equation}
   \eps(\theta) = \alpha P_{l}(\cos\theta)
\end{equation}
and there is no additional scaling factor, unlike in the case of the cylinder, where the mode wavelength had to be introduced.
In the spherical coordinate system\footnote{We use the same letter, $r$, for the radial coordinate in the cylindrical and spherical systems. The correct meaning is determined by the context in which it is used.} $r,\theta,\phi$,
 the unperturbed potential is 
\begin{multline}
   \Phi({r}) =   \\ 
   2\pi \Delta \rho   \left\{
    \begin{array}{ll}
 \frac{1}{3} r^2 (w-1)- R^2 (w^{1/3}-1) & 0\leq r\leq R \\
 \frac{1}{3} r^2 w + \frac{2 R^3}{3 r}- R^2 w^{1/3} & R<r\leq R_c~,
\end{array}
\right. 
\end{multline}
where the cell radius is   $R_c = R/w^{1/3}$. 
For the spherical cell, the Bond number is
\begin{equation}
   \zeta = \frac{4 \pi   \Delta \rho ^2 R^3 (w-1)^2}{9 \sigma}
\end{equation}
and the Green's function takes the form
\begin{equation}
    G_{sph}(\vbx,\vbx')   =       \sum_{l=0}^\infty g_l(r,r') P_l(\cos \gamma)~,
   \label{greensph}
\end{equation}
where $P_l$ are the Legendre polynomials, $\gamma$ is the angle between the points $\vbx$ and $\vbx'$, and
is given by $\cos\gamma = \cos\theta\cos\theta' + \sin\theta\sin\theta'\cos(\phi-\phi')$ 
\cite{jackson}. As was previously discussed, the radial functions $g_l(r,r')$ depend on the boundary conditions imposed on the cell surface and for the generalized Green’s function these are 
 \begin{equation}
   g_l^{\rm gen}(r,r') =
   \left\{ 
   \begin{array}{ll}
   \fracd{1}{r_>}+\fracd{r^2+r'^2}{2 R_c^3}-\fracd{9}{5 R_c} & ,~~l=0 \vspace{.7em} \\ 
   \fracd{r_<^l}{r_>^{l+1}} \left[\fracd{l+1}{l} \; \left(\fracd{r_>}{R_c}\right)^{2l+1} + 1\right] & ,~~l>0~,
         \end{array}
    \right.
 \end{equation}
where $R_c$ is now the spherical cell radius. The stability function defined by the expression $f = 25\, \delta^2\tilde{E}/(4\pi ^2 \Delta \rho ^2 R^3 \alpha ^2)$ is now a function of only the volume fraction, $w$, and the Bond number, $\zeta$. Below we present the explicit form for the three lowest deformation modes:
 \begin{equation}
f(w,\zeta) =
   \left\{ 
   \begin{array}{ll}
   \frac{1}{3} \left(9 w^{5/3}+10 w-4\right)+\frac{40 }{9 \zeta }(w-1)^2 & ~~l=2 \\
   \frac{50}{147} \left(4 w^{7/3}+7 w-4\right)+\frac{500 }{63 \zeta }(w-1)^2 & ~~l=3 \\
   \frac{25}{162} \left(5 w^3+12 w-8\right)+\frac{100}{9 \zeta } (w-1)^2 & ~~l=4~. 
         \end{array}
    \right.
 \end{equation}
In Fig.~\ref{ws-stabmap-sph}, the instability regions for the different modes are shown. As may have been expected for $w\to 0$,
the quadrupole mode recovers the well-known Bohr--Wheeler instability for a nucleus in the liquid drop model \cite{Bohr:1939ej}. The Bohr--Wheeler condition corresponds to a Bond number 
\[\zeta= \frac{10}{3}~ . \]
If the virial theorem is applied, the relation between the surface tension and the charge contrast is 
\begin{equation}
   \sigma =\frac{4}{15}\pi   \Delta \rho ^2 R^3(w - 3 w^{1/3} + 2)
   \end{equation}
and again similar to the cylindrical W--S cell, the Bond number becomes bounded. For the spherical W--S cell, the value of the Bond number takes the form
\begin{equation}
{\rm Bo_e^{vir}} = \zeta(w) = \frac{5 (w-1)^2}{3 (w-3 w^{1/3}+2)} ~.
\end{equation}
The relation is shown in Fig.~\ref{Bovir-plot} and the values of the Bond number are confined to the range
 $\frac{5}{6} < \rm Bo_e^{vir} < 5$.
The restriction of the Bond number values makes the stability function $f^{\rm vir}(w)$ positive for any modes and  
any values of the volume fraction, $w$. The spherical charged ball in a spherical W--S cell is always stable.
\section{Conclusions}
The stability of pasta phases in the cylindrical and spherical Wigner--Seitz cell approximation was analyzed. The stability of a given 
mode was determined by inspecting the second-order energy variation with respect to the proton cluster shape perturbation.
As an illustration, we have also examined the charged rod with finite surface tension placed in vacuum. This case could be treated as a 
limiting case when the rod placed in a cylindrical W--S cell becomes very thin. In the cylindrical case, different perturbation modes are 
unstable in different regions of parameter space. The relevant parameters are the following: the mode wavelength and the electric Bond number, which is the measure of the magnitude of the electric forces. 
Due to the virial theorem, the Bond number is no longer a free parameter and is strictly related to the ratio of the
 proton cluster volume to the cell volume, $w$. Because of that, the Bond number never grows to large values, electric forces are reduced, and most of the modes cannot grow. The only unstable mode is the Plateau—Rayleigh mode ,which is unstable for very small values of the volume fraction, $w$.
These two cases, vacuum and in the W--S cell, lead to opposite conclusions about cluster stability, which have their roots in the boundary conditions.
 
 Similar considerations were also conducted for the spherical W--S cell. 
Again, the virial theorem constrains the value of the electric Bond 
number and completely stabilizes the charge ball inside the W--S cell. 
The only instability concerns the quadrupole perturbation, in the limit of $w \to 0$ that corresponds to the Bohr--Wheeler fission of an isolated nucleus. In general, one may conclude that the strong boundary 
conditions imposed for the W--S cell, i.e., an electric field that vanishes on the cell boundary, stabilize the possible deformations. A more realistic picture will be obtained when the true periodicity of the system is introduced.

\begin{acknowledgments}
One of the authors (SK) would like to acknowledge Stefan Typel
who encouraged us to research the pasta stability issue.
\end{acknowledgments}

\end{document}